\documentclass[%
reprint,
superscriptaddress,
 amsmath,amssymb,
 aps,
longbibliography,
floatfix,
]{revtex4-2}

\usepackage[dvipsnames,svgnames,x11names,hyperref]{xcolor}
\usepackage{siunitx}
\usepackage{tabularx}
\usepackage{color}
\usepackage{ulem}
\usepackage{graphicx}
\usepackage{dcolumn}
\usepackage{bm}
\usepackage{hyperref}
\usepackage[mathlines]{lineno}
\usepackage{mathrsfs}
\usepackage{braket}
\usepackage{placeins} 
\usepackage{float} 

\hypersetup{colorlinks=true, 
	linkcolor={blue!75!black!80!yellow},
	citecolor={blue!75!black!80!yellow}, 
	urlcolor=black 
	}
\makeatletter \renewcommand\@make@capt@title[2]{%
\@ifx@empty\float@link{\@firstofone}{\expandafter\href\expandafter{\float@link}}%
\sffamily{\textbf{#1}}\@caption@fignum@sep#2 }

\begin{document}

\title{A roadmap toward the theory of vibrational polariton chemistry}

\author{Derek S. Wang}
\email{derekwang@g.harvard.edu}
\affiliation{Harvard John A. Paulson School of Engineering and Applied Sciences, Harvard University, Cambridge, MA 02138, USA}

\author{Susanne F. Yelin}
\affiliation{Department of Physics, Harvard University, Cambridge, MA 02138, USA}

\begin{abstract}
\noindent The field of vibrational polariton chemistry was firmly established in 2016 when a chemical reaction rate at room temperature was modified within a resonantly tuned infrared cavity without externally driving the system. Despite intense efforts by scientists around the world to understand why the reaction rate changes, no convincing theoretical explanation exists. In this perspective, first, we briefly review this seminal experiment, as well as relevant experiments that have since followed that have hinted at the roles of reactant concentration, cavity frequency, and symmetry. Then, we analyze the relevance of leading theories, such as quantum electrodynamics-modified transition rate theories, the photonic solvent cage effect, the impact of dissipation from dark states, bond strengthening \textit{via} intramolecular vibrational energy redistribution, and collectively enhanced local molecular properties. Finally, we construct a roadmap toward the theory of vibrational polariton chemistry by suggesting experiments to test theories and new paths for theorists. We believe that understanding the importance of the onset of the strong coupling regime, designing experiments to capture changes in reaction pathways, and further developing the theories of cavity-modified intramolecular vibrational energy redistribution and collectively enhanced local molecular properties are crucial next steps. We hope this perspective will be a valuable resource for guiding research in the field of vibrational polariton chemistry.
\end{abstract}

\date{\today}

\maketitle

\section{Introduction} \label{sec:intro}
\begin{figure}[tbhp]
\centering
\includegraphics[width=0.8\linewidth]{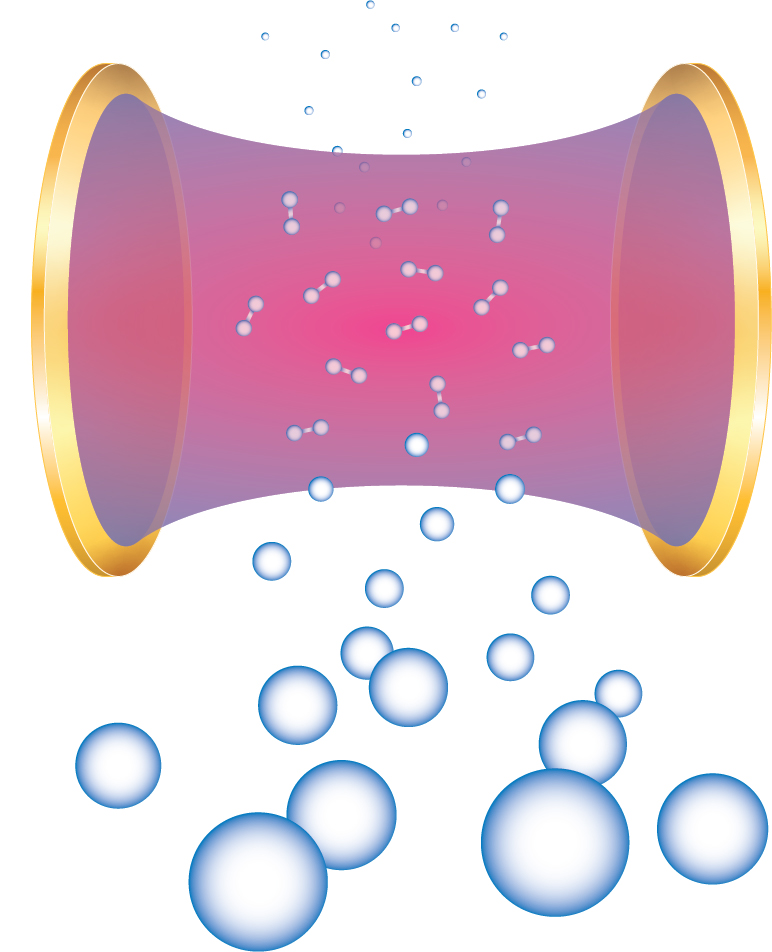}
\caption{Schematic of a model cavity-modified chemical reaction. When the cavity mode is resonant with a vibrational excitation of the molecule, dissociation of the molecule is slowed down.
}
\label{fig:schematic}
\end{figure}

Chemists have long fantasized about influencing chemical reactivity, creating molecules impossible with the conventional toolset of synthetic chemistry, and unraveling mechanisms of energy transfer relevant to photosynthesis by selectively exciting a vibrational mode. Studies in the 1980s and 1990s \cite{Frei1980, Frei1983, Frei1983a, Bucksbaum1990}, however, found that intramolecular vibrational energy redistribution (IVR) severely limited the efficiency of mode-selective excitation unless the reactions took place in cryogenic temperatures. In 2016, Thomas \textit{et al.} revitalized interest in this pursuit and established the field of vibrational polariton chemistry when they demonstrated that a chemical reaction, at room temperature, could be modified within the environment of a resonantly-tuned infrared cavity without externally driving the system \cite{Thomas2016}, as shown schematically in Fig. \ref{fig:schematic}. This discovery opened the door to myriad tantalizing possibilities, but despite intense efforts since by scientists around the world to understand why the reaction rate changes, at the time of writing, no convincing explanation exists. 

In this perspective geared at both those interested in joining the field and veterans who would appreciate a guide through the slew of theoretical papers that have been released in the past five years, we evaluate theories put forth thus far against the experimental evidence amassed and suggest future studies toward the theory of vibrational polariton chemistry. First, we briefly review the seminal experiment from 2016, as well as relevant experiments that have since followed that have hinted at the roles of symmetry, concentration, kinetics, and strong coupling. We focus on experimental details and results that a robust theory of vibrational polariton chemistry must include and justify, respectively. Then, we discuss the major theories, focusing on their relations to the well-established theories of chemical reaction dynamics and analyzing their relevance to the conditions achieved in the experiments. Finally, we construct a roadmap toward the theory of vibrational polariton chemistry by suggesting experiments to test theories and new paths for theorists. Based on our evaluation of the theories and experiments, we believe that studying the roles of symmetries and cavity losses in intra- and inter-molecular energy transfer, as well as extending theories concerning collectively enhanced-modifications to local environments to chemical reactions, are crucial next steps for the polariton chemistry community. We suggest specific studies for re-visiting old experiments with new insights and for teasing out the role of intra- and inter-molecular energy transfer. We also look toward the future of vibrational polariton chemistry once the first batch of experiences have been understood with suggestions for future applications.

\section{Experiments} \label{sec:experiment}

\begin{figure}[tbhp]
\centering
\includegraphics[width=1.0\linewidth]{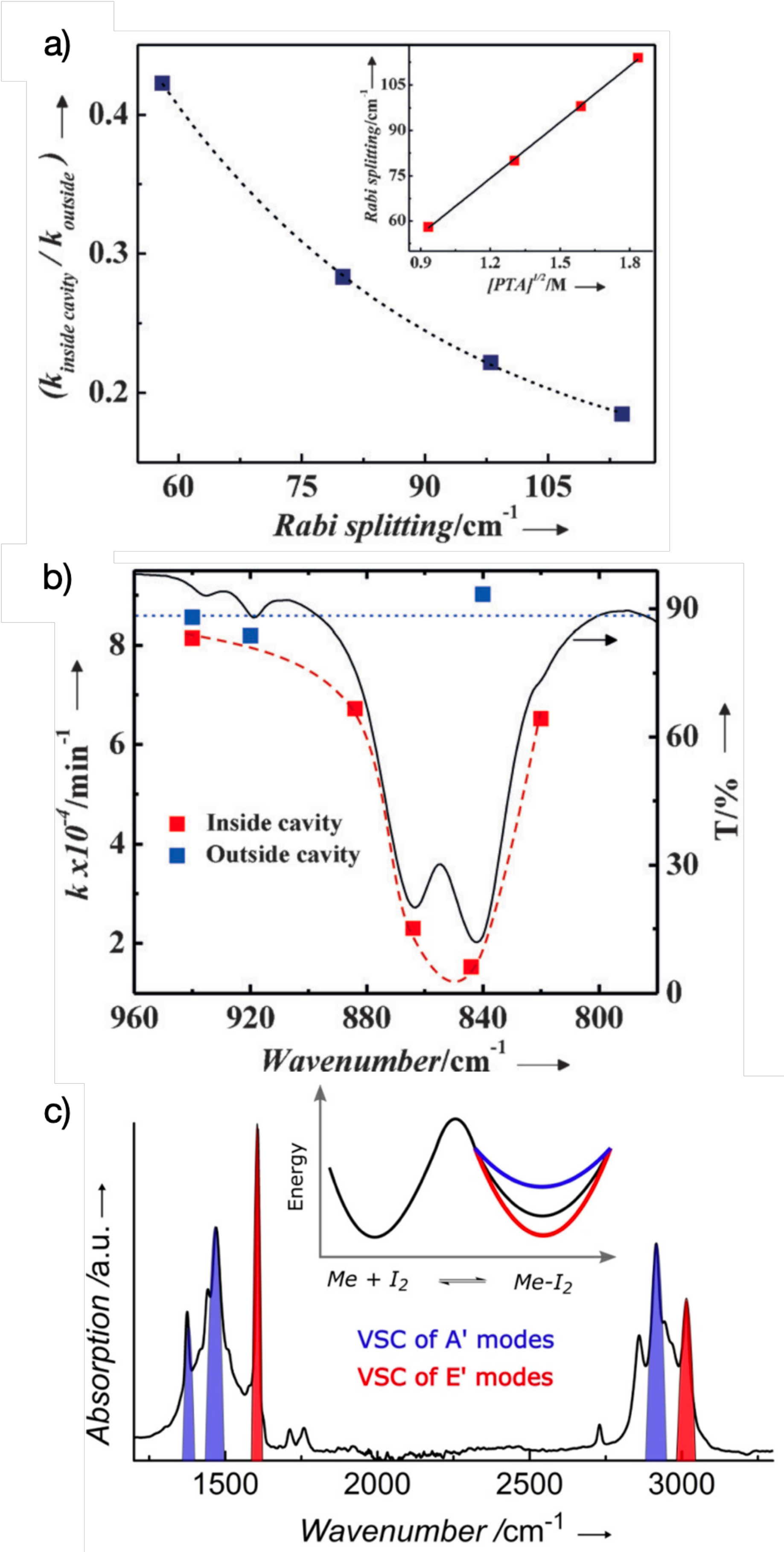}
\caption{Experimental data demonstrating the \textbf{(a)} $N$-dependence, \textbf{(b)} resonance, and \textbf{(c)} symmetry effects. (a) and (b) are reproduced from Ref. \citenum{Thomas2016}, and (c) is reproduced from Ref. \citenum{Pang2020}, all with permission.
}
\label{fig:experiments}
\end{figure}

We begin by reviewing the seminal experiment by Thomas \textit{et al.} in Ref. \citenum{Thomas2016}, representative of further experiments confirming similar trends \cite{Hirai2020}. We favor a more detailed description of the experimental conditions, as we will later show in detail which experimental conditions the various proposed theories do or do not correspond to. In this work, they study the rate of a simple reaction inside an infrared (IR) cavity without externally driving the system with IR photons. 
The rate-limiting step is thought to involve the attack of the fluoride ion (F$^-$) from tetra-$n$-butylammonium fluoride (TBAF) onto the silicon atom of 1-phenyl-2-trimethylsilylacetylene (PTA), resulting in dissociation of PTA into two products upon breaking the Si-C bond. The entire reaction occurs in a polar solvent, methanol. The PTA reactant has several well defined vibrational modes that can be identified in IR spectroscopy, including a mode at 860 cm$^{-1}$ with a full-width at half-maximum of 39 cm$^{-1}$ that largely corresponds to local vibration of the Si-C bond broken during the course of the reaction. The cavity consists of two parallel mirrors separated by a few microns, the precise distance of which can be tuned to shift the resonance frequencies given as $10^4 m / (2 n L)$ in wavenumber units cm$^{-1}$, where $m$ is the integer mode order, $n$ is the refraction index, and $L$ is the length of the cavity in $\mathrm{\mu}$. Importantly, the cavity mirrors are gold with several hundred nanometers of silica deposited on the inner surfaces, such that neither the gold itself nor its plasmonic near fields directly contact the reaction mixture. To compare the reaction inside versus outside the IR cavity, they also create an identical device, except with no gold deposited such that the vacuum electric field cannot be concentrated within the volume of the device. Because the reactant and product have different refractive indices $n$, the progress of the reaction can be tracked by measuring the frequency drift of one of the higher order cavity resonance peaks that are far off-resonant from any of the vibrational modes of any of the molecules involved in the reaction. They tune the $m=2$ mode of the cavity with a FWHM of 30 cm$^{-1}$ in resonance with the vibrational mode corresponding largely to Si-C vibration in PTA at 860 cm$^{-1}$, resulting in a Rabi splitting $\hbar\Omega_{\mathrm{R}}$ of 98 cm$^{-1}$ and putting the system squarely within the collective strong-coupling regime. Interestingly, they show that the reaction rate constant decreases inside the cavity versus outside the cavity.

To better understand this decrease in reaction rate, Thomas \textit{et al.} sweep three experimental conditions in the aforementioned setup. First, as shown in Fig. \ref{fig:experiments}(a), they sweep the concentration of PTA in the solution from 0.87 M to 3.37 M. They then show there is a linear relationship between the square root of the concentration of PTA and the Rabi splitting observed in the IR transmission spectrum and a nonlinear decrease of the relative reaction rate constant inside the cavity versus outside the cavity, from $\sim$0.4 to $\sim$0.2, for increasing Rabi splitting. Second, as shown in Fig. \ref{fig:experiments}(b), they sweep the frequency of the second cavity mode from 820 to 940 cm$^{-1}$ and show that the reaction rate constant decreases the most around 860 cm$^{-1}$, while the reaction outside the cavity shows no dependence on the distance between the dielectric mirrors (``outside the cavity") that corresponds to the resonant frequencies of the cavity with gold mirrors. Notably, the linewidth of the dip in reaction rate inside the cavity closely follows the linewidth of the dip in IR transmission of PTA outside the cavity. Third, they sweep the temperature from 294 to 312 K for the reaction both inside and outside the cavity and show a generally linear relationship between the logarithm of the reaction rate constant vs. the inverse temperature for both conditions. These curves are fit to the Eyring equation of chemical reaction dynamics, which looks similar to the empirically derived Arrhenius equation but is in fact derived from statistical mechanical considerations:
\begin{equation}
    k = \frac{\kappa k_\mathrm{B} T}{h}\mathrm{e}^{\frac{\Delta G^\ddag}{RT}},
\end{equation}
where $k$ is the reaction rate constant, $\kappa$ is the transmission coefficient that quantifies how often reactants successfully pass through the activation barrier to the product side without returning and is generally assumed to be 1, $T$ is the temperature, the free energy of activation is $\Delta G^\ddag = \Delta H^\ddag - T\Delta S^\ddag$, the activation enthalpy is $\Delta H^\ddag$ and activation entropy is $\Delta S^\ddag$. The Eyring equation, therefore, parametrizes a reaction by just two parameters, the activation enthalpy $\Delta H^\ddag$ and activation entropy $\Delta S^\ddag$. For [PTA]=2.53 M and $\hbar\Omega_{R}=98$ cm$^{-1}$, Thomas \textit{et al.} show that $\Delta H^\ddag$ and $\Delta S^\ddag$ both increase inside the cavity, from 39 to 96 kJ/mol for the former and from -171 to 7.4 J/(K mol) for the latter.

Based on the first and second sweeps, Thomas \textit{et al.} claim that vibrational strong coupling of the Si-C vibration with the cavity mode is directly responsible for the changes in reaction rate constant. Based on the third sweep, they speculate that the reaction mechanism may have changed from associative, where the initial step is F$^{-}$ attack on the silicon atom to form an intermediate with pentavalent coordination, to dissociative, where Si-C bond breaks before the F$^-$ attacks. 

It is still unclear whether onset of the strong coupling regime, where the light-matter coupling rate is larger than the cavity and emitter loss rates, is necessary for the observed changes to the chemical reactivity, as opposed to, say, coupling to a cavity mode with a vacuum electric field or light-matter coupling rate above some threshold. Whether the mechanism of a reaction changes in a cavity or whether the original pathway is simply accelerated or decelerated is also still an open question. We discuss steps to resolve both questions in Section \ref{sec:outlook}. Nonetheless, we can be sure of at least the following characteristics of vibrational polariton chemistry: 1) The $N$-dependence: The reaction rate changes as the concentration, or number of particles $N$ in the volume of the IR cavity, increases, and 2) the resonance effect: There are particular IR cavity frequencies at which the effect is the strongest.

In two recent follow-up papers by the Ebbesen group \cite{Pang2020, Sau2021}, they demonstrate what seems to be another critical feature of vibrational polariton chemistry: the symmetry effect. For brevity purposes, here we discuss only the 2020 paper \cite{Pang2020} that first introduces this idea,
where Pang \textit{et al.} study the charge transfer complexation between mesitylene and iodine inside an IR cavity constructed very similarly to the one in the seminal experiment. This reaction is notable as it explicitly involves multiple electronic potential energy surfaces, suggesting that non-adiabatic couplings between them are crucial, whereas previous studies likely occur entirely on the electronic ground state potential energy surface.

They first couple a vibrational mode of mesitylene with E' symmetry to the cavity and measure the equilibrium concentration of the charge transfer complex in the UV/vis region as a function of the initial concentration of mesitylene. From this curve, they extract two parameters: the equilibrium constant $K_\mathrm{DA}$, where D and A stand for electron donor and acceptor, respectively, and the absorption coefficient $\epsilon_\mathrm{DA}$. Note that $K_\mathrm{DA}$ relates to the change in free energy $\Delta G ^\circ$ between the reactant and product, not the free energy activation barrier, as measured in the seminal experiment. The absorption coefficient $\epsilon_\mathrm{DA}$, meanwhile, is dependent on the exact geometry of the mesitylene-iodine complex formed. Fascinatingly, they show that both $K_\mathrm{DA}$ and $\epsilon_\mathrm{DA}$ can change at the onset of the strong-coupling regime when the Rabi splitting is large enough to produce two visible polaritonic peaks in the IR transmission spectrum, providing some evidence, albeit not conclusive, that strong coupling is necessary for cavity-modified chemistry. The change in $K_\mathrm{DA}$ corresponds to a change in $\Delta G^\circ \gg$ both $k_\mathrm{B}T$ and $\hbar\Omega_\mathrm{R}$. As shown in Fig. \ref{fig:experiments}(c), they then sweep the cavity frequency across the range of IR vibrational frequencies of the reactants and observe the symmetry effect: strongly coupling to vibrational modes with A' symmetry decreases $K_\mathrm{DA}$ and increases $\epsilon_\mathrm{DA}$ relative to those outside the cavity, while coupling to E' modes increases $K_\mathrm{DA}$ and does not drastically affect $\epsilon_\mathrm{DA}$. Furthermore, further increasing the Rabi splitting does not change these values, suggesting that upon reaching the strong coupling regime, this effect is dominated purely by the symmetry of the vibrational modes. 

To see whether this correlation between vibrational mode symmetry and change in equilibrium properties of the complex can be extended to causation, they study deuterated mesitylene and replace mesitylene with benzene. The molecular vibrations of deuterated mesitylene with identical symmetries are at shifted frequencies due to the isotope effect, and they show that strongly coupling to these frequency-shifted modes results in identical changes to $K_\mathrm{DA}$ and $\epsilon_\mathrm{DA}$ compared to non-deuterated mesitylene. In addition, by shifting the molecular vibrational frequencies, they are able to couple uniquely to the solvent vibrational modes and show no change. By replacing mesitylene with benzene, they study the influence of mode symmetry in a different molecule that possess only E modes and observe similar changes to $K_\mathrm{DA}$ and $\epsilon_\mathrm{DA}$ compared to coupling to E' modes in mesitylene. Overall, this study provides evidence for the symmetry effect, where the symmetry of the vibrational mode coupled to the cavity is an important factor for observing coupling-dependent chemical reactivity.

\section{Theories} \label{sec:theories}

\begin{figure*}[tbhp]
\centering
\includegraphics[width=1.0\linewidth]{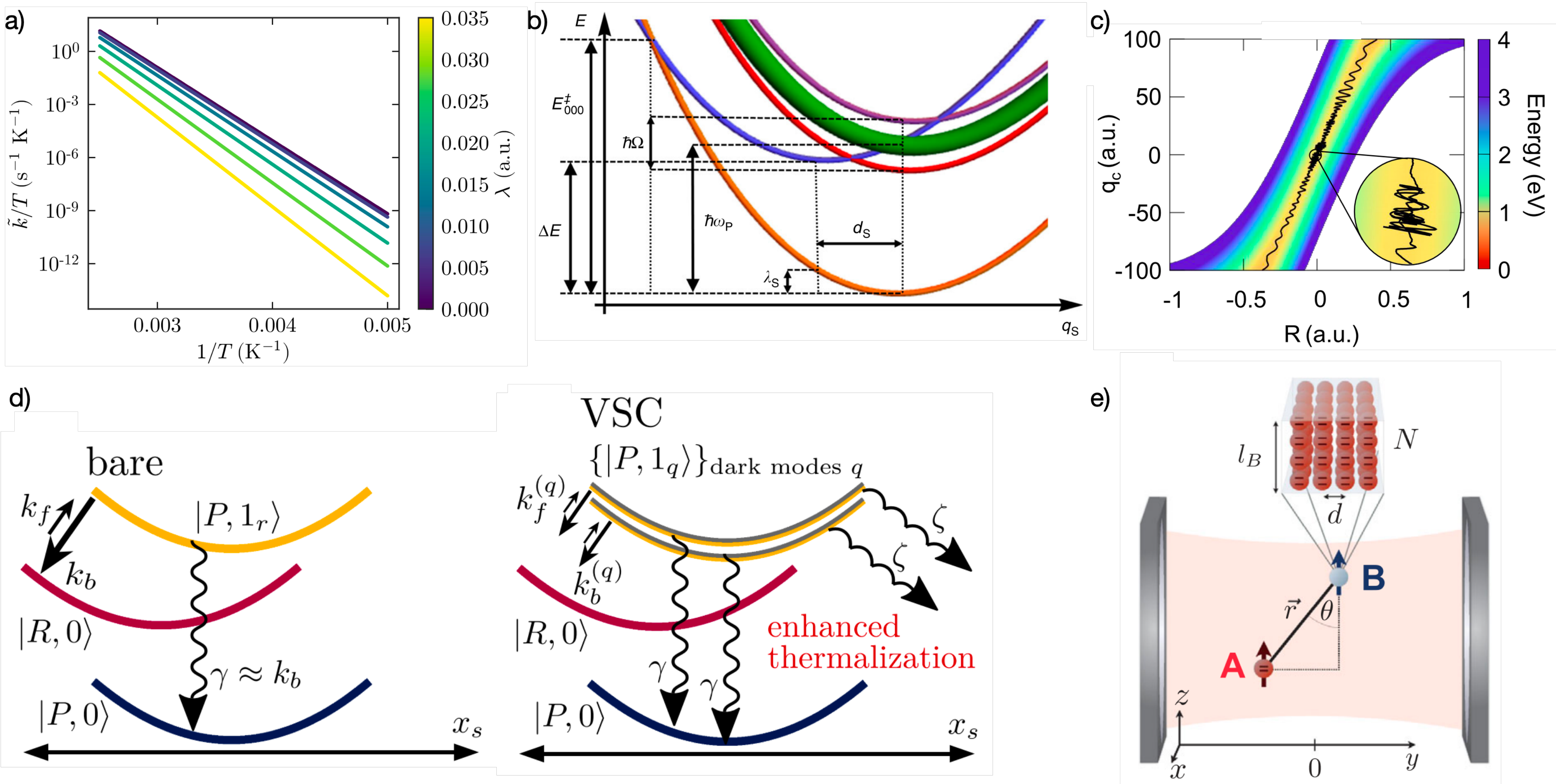}
\caption{Theories. \textbf{(a)} Increasing the light-matter coupling strength up to $\lambda=0.035$ corresponding to a Rabi splitting that is 0.1 of the bare transition energy, coupling strengths inaccessible in current Fabry-Perot cavities, shifts the Eyring curve. \textbf{(b)} Potential energy surfaces demonstrating that the activation energy decreases from the reactant (blue) to the lower polariton (red) of the product, although transfer to the dark states (green) are still preferred. Thermal relaxation de-excites the lower polariton, dark states, and upper polariton (purple) to the vibrational round state of the product (orange). \textbf{(c)} The photonic mode $q_c$ acts as a solvent cage along the reaction coordinate $R$. 
\textbf{(d)} Outside the cavity, the reaction rate is a competition between the forward rate $k_f$ from the reactant $|R, 0\rangle$ to the vibrationally excited product $|P, 1_r\rangle$, backwards rate $k_b$, and thermalization rate $\gamma$. Inside the cavity, the dark modes enhance the thermalization rate, overall accelerating the reaction rate. \textbf{(e)} An emitter A coupled to an ensemble of emitters B can experience an effective increase in coupling to the cavity. Figures are reproduced from Refs. \citenum{Galego2019,Campos-Gonzalez-Angulo2019, Li2021a, Du2021, Schutz2020}, respectively. 
}
\label{fig:theories}
\end{figure*}

From the experiments conducted thus far,  we note the following three robust features of vibrational polariton chemistry: 
1) the $N$-dependence, where the reaction rate changes with the number of reactant molecules $N$ in the cavity volume, 2) the resonance effect, where only certain cavity frequencies change the reaction rate, and 3) the symmetry effect, where effects are seen when coupling the cavity to vibrational modes with particular symmetries. Therefore, the theory of vibrational polariton chemistry should manifest at least these three effects. Other experimental conditions that should be taken into account include room temperature, the presence of solvent, disordered geometrical configurations of reactants, inter- and intramolecular interactions including other vibrations and electronic transitions, and the spatially-dependent cavity spectral profile, among other system complexities. Keeping these considerations in mind, we turn to the major theoretical developments put forth in the five years since the seminal experiment, starting first with analyses based on intuition from standard quantum optics models that have worked with some success in explaining electronic polariton chemistry. For each theory, we discuss the basic foundation and results, in which ways they do or do not match experimental observations, and how they can be built upon in the near future toward a robust theory of vibrational polariton chemistry.

A common (but, as we will discuss, misguided) argument to convey an intuitive understanding of how vibrational polariton chemistry works might go as follows: Consider a simple two-level system, where the ground state corresponds to a reactant in its electronic and vibrational ground state, and the excited state corresponds to a reactant still in its electronic ground state but in a vibrationally excited state. This simple two-level system interacts with a resonant, unpopulated cavity mode. This interaction can be represented by the Jaynes-Cummings Hamiltonian that assumes the dipole approximation (where the electric field of the cavity mode is constant over the volume of the transition dipole density), the rotating wave approximation (where counter-rotating terms are dropped), and the dipole self-energy term is dropped. The light-matter interaction between the vibrational transition dipole moment and vacuum electric field of the cavity results in mixing, such that the bare vibrational excitation and cavity mode are no longer eigenstates of the system. Instead, the excited eigenstates become a lower and an upper polariton split symmetrically about the bare energies by the Rabi splitting. The argument is then that, because the upper polariton contains a contribution from the vibrational excitation and is higher energy, the vibrational energy gets a ``free" energy boost, enabling the molecule to hop the activation energy barrier more easily during a reaction, in accordance with the Eyring equation. When there are $N$ molecules, the Jaynes-Cummings Hamiltonian becomes the Tavis-Cummings Hamiltonian, and it can be easily shown that the Rabi splitting between the lower and upper polaritons is enhanced by a factor proportional to $\sqrt{N}$, supposedly increasing the magnitude of the ``free" energy boost and further enabling the molecule to tunnel through the activate energy barrier. It is unclear, however, whether there is truly a ``free" energy boost in the single two-level system case; in fact, if this mechanism were correct, the activation energy barrier should change linearly with the Rabi splitting, but in fact, the effective free energy of activate changes nonlinearly \cite{Thomas2020}. In addition, it is unclear whether, in the $N$ two-level systems case, this energy boost can be localized onto a single molecule, especially when one considers that the eigenstates include not only a lower and upper polariton, but also $N-1$ dark states that are totally decoupled from the cavity mode with properties similar to those of bare excitations.

In Ref. \citenum{Galego2019}, Galego \textit{et al.} rigorously investigate this intuitive theory. They study a simple Shin-Metiu model that comprises one electron and three nuclei in one dimension, where two nuclei are fixed in place, while the electron and other nucleus interact with the other nuclei with a normal Coulomb and softened Coulomb potential, respectively. While the Shin-Metiu can easily be parametrized to bring electronic excited state potential energy surfaces close to each other and to the ground-state and, thus, demand inclusion of non-adiabatic terms, in this case, they parametrize the system such that only the electronic ground state is relevant. The Shin-Metiu ``molecule" is placed in a cavity, where the dipole self-energy term in the light-matter interaction Hamltonian has been dropped and the vacuum electric field is allowed to be spatially varying, \textit{i.e.} not the dipole approximation. They compute the full quantum reaction rate of the electron and freely moving nucleus hopping from potential well to the other using the formalism of Miller \textit{et al.} originally described in Refs. \citenum{Miller1974, Miller1983} and summarized nicely in Ref. \citenum{Manolopoulous2008}. On a single-molecule level, they show that the reaction rate can indeed be shifted, as shown in Fig. \ref{fig:theories}(a), and that the ground state structure of the molecule can be modified, but the coupling strengths required are likely only achievable inside nanoscopic cavities realized by, for instance, the mode volume of a spacer between a nanoparticle and a mirror that is many orders of magnitude smaller than the mode volume inside the IR cavities studied experimentally. In addition, they observe no resonance effect. Generalizing their approach to $N$ molecules, they observe also no collective effects, unless the molecules are aligned along the same direction. Preferential alignment of the molecules is naively unlikely in the micron-scale cavities of the recent experimental demonstrations, although recent cavity molecular dynamics simulation of water molecules under vibrational ultrastrong coupling hint at dynamic oriental preferences of molecules inside such cavities \cite{Li2020}. All these conclusions based on transition rate theory are largely supported by other groups, including T. E. Li \textit{et al.} in Ref. \citenum{Li2020a} who take a standard chemical rate theory approach where the cavity and nuclear modes are treated classically. Therefore, it seems that the ``intuitive" explanation of Rabi-boosted polaritons more easily hopping across activation energy barriers does not pass the muster of this more rigorous theoretical approach.

Campos-Gonzalez-Angulo \textit{et al.} \cite{Campos-Gonzalez-Angulo2019} propose a cavity-modified Marcus-Levich-Jortner model with a mechanism of modulating chemical reaction rates similar to the intuitive argument given before. This model predicts reaction rates involving electron transfer from a donor to an acceptor across nuclear coordination-dependent potential energy surfaces. In a cavity resonantly tuned with a molecular vibration of $N$ products, the authors show that electron transfer to the $N-1$ dark states near the energy of the bare excitations is entropically favored, as in Fig. \ref{fig:theories}(b). However, they claim that electron transfer to the lower polariton, shifted downwards in energy by the Rabi splitting, in fact dominates reaction kinetics. This mechanism satisfies both the resonance and $N$-dependence requirements of a theory of vibrational polariton chemistry because the Rabi splitting is maximized when the cavity is resonant with the vibration of interest and is proportional to $\sqrt{N}$. However, because this theory involves a excited electronic potential energy surface with non-adiabatic coupling to the ground-state potential energy surface, this theory is likely not suitable for the seminal experiment, although in this sense, it may be relevant to Ref. \citenum{Pang2020}, where they study the symmetry effect in an electron transfer complexation reaction. Recently, however, in Ref. \citenum{Vurgaftman2020}, Vurgaftman \textit{et al.} dispute this mechanism, claiming that when the dispersive nature of the cavity is included, the density of states at the frequency of the lower polariton is in fact dominated by dark states unless the Rabi splitting is more than an order of magnitude larger than the linewidths of the bare excitations, a far stricter regime not applied in experiments.

X. Li \textit{et al.} in Ref. \citenum{Li2021a, Li2021} question a crucial, oft-taken assumption of the transition rate theories: instead of assuming the transmission coefficient $\kappa=1$, they rigorously determine its dependence on the coupling between the vibrational modes of a molecule and the cavity. As Galego \textit{et al.} do in Ref. \citenum{Galego2019}, they study the Shin-Metiu model with and without coupling the vibrational mode of interest to other vibrational modes in the molecule. From Grote-Hynes theory \cite{Grote1980, Hanggi1990}, they derive the condition under which $\kappa$ is at a minimum: when $\kappa$ equals the barrier frequency $\omega_b$. $\omega_b^2$ is proportional to the curvature of the reaction barrier when the light-matter coupling strength is low and decreases as the light-matter coupling strength increases. This calculated change to $\kappa$ does not change the free energy of activation $\Delta G^\ddag$, but it can result in an effective change on the order of a few kJ/mol, a non-trivial amount that leads to rate slow-downs. Physically, resonantly tuning the cavity frequency close to the barrier frequency results in the photon mode acting like a solvent degree of freedom, trapping the molecule near the transition state and ultimately slowing the reaction rate, as in Fig. \ref{fig:theories}(c). While this photonic solvent cage effect does have a resonant cavity frequency, the value of the barrier frequency calculated by Sch\"afer \textit{et al.} in Ref. \citenum{Schafer2021} of 74-94 cm$^{-1}$ does not correspond to the resonance frequency observed in the seminal experiment of 860 cm$^{-1}$. Finally, the connection between this theory and realizing the $N$-dependence is also unclear. X. Li \textit{et al.} compensate for the lack of $N$ molecules in their model by increasing the vacuum electric field to realize Rabi splittings of the same order of magnitude as in the seminal experiment, but the mapping between $N$ molecules and a weaker vacuum electric field strength versus one molecule and a stronger vacuum electric field strength is exact only under the approximations of the Tavis-Cummings model that, among many approximations, neglects the complex, nuclear configuration-dependent energy structure of the molecules by assuming that they are just two-level systems, as well as drops counter-rotating terms and assumes the dipole-approximation. As will be discussed later, however, theories have emerged that attempt to explain how local changes can be enhanced by a collectively coupled ensemble and may account for the $N$-dependence in the photonic solvent cage theory as well as others lacking a direct connection to the $N$-dependence.

Sch\"afer \textit{et al.} in Ref. \citenum{Schafer2021} use quantum electrodynamical density functional (QEDFT) \cite{tokatly2013, ruggenthaler2014, flick2019lmrnqe, flickexcited, Wang2020LossQEDFT}e to run real-time simulations of the rate-limiting step involving F$^-$ ion attacking the Si-C bond in PTA and report, for the first time, microscopic evidence of the resonance effect. QEDFT is a first-principles framework for fully describing the interactions between electrons, nuclei, and photonic degrees of freedom, where the the dipole self-energy and counterrotating terms are kept and the dipole approxiamtion is taken. In their simulations, they perform 30 trajectories where the reactant molecules are launched toward each other with initial velocities sampled from a thermal distribution of 300 K, which is within the range of temperatures studied by Thomas \textit{et al.} in Ref. \cite{Thomas2016}. The molecules are coupled to a single, lossless photon mode tuned under the dipole approximation either in resonance (860 cm$^{-1}$) or out of resonance (425 cm$^{-1}$) with the Si-C bond at 860 cm$^{-1}$. Just as X. Li \textit{et al.} do in their theoretical studies on a single molecule, they choose the vacuum electric field to be strong enough to result in Rabi splittings approximately equal to those in the seminal experiment, which again naturally lead to questions about how the observed change in reaction rate changes as the reactant concentration increases. Averaging over the trajectories, they find that when the cavity is resonantly tuned to the Si-C vibration, the reaction trajectory is trapped in the local minimum corresponding to the pentavalent intermediate PTA-F$^-$ longer than when the cavity is not resonantly tuned to the Si-C vibration. To understand why the reaction trajectory gets trapped, they plot the difference in vibrational mode occupations between the resonant and off-resonant conditions and show that, over the course of the reaction, the resonant cavity mode better enables energy transfer between the vibrational mode corresponding to Si-C vibration and other vibrations, especially those involving the F-Si-C-C chain. 

While these initial results are promising, there remain some difficulties in comparing the computational results for trajectories of single molecules with reaction rate constants from the seminal experiment. For instance, as in the seminal experiment, they sweep the cavity frequencies but in a wider range, from 400 to 1700 cm$^{-1}$, and show that the time-averaged Si-C distances are longer inside the cavity than outside the cavity, \textit{i.e.} when the light-matter coupling strength is zero, for cavity frequencies between 500 cm$^{-1}$ and 1700 cm$^{-1}$, whereas in the seminal experiment, the reaction rate constant inside the cavity approaches the rate outside the cavity at just $\sim$800 to 960 cm $^{-1}$. Clearly, the FWHM of the change in Si-C bond distance is much larger in the simulation than the reaction rate difference is in the experiment, but whether differences in fact result in the same observed cavity frequency-dependent effect is unclear. Furthermore, they do not calculate the time-averaged Si-C distance or analogous metric of reaction progress for varying coupling strength nor the temperature, as Thomas \textit{et al.} do in the first and third sweeps of Ref. \citenum{Thomas2016}. Finally, as in the previous studies, no formal connection to the $N$-dependence is given. However, the authors do note that recent works where collective strong-coupling can lead to strong coupling on the local level as individual molecules experience collectively enhanced dipole moments. 

Many of these aforementioned theories struggle to explain both the resonance effect and $N$-dependence, but perhaps these two effects do not need to be inextricably tied together by the same physical mechanism. In Ref. \citenum{Du2021}, Du and Yuen-Zhou propose an explanation for the $N$-dependence based on the enhancement of vibrational dissipation \textit{via} dark states when the $N$ molecular vibrational modes are disordered. First, they study a chemical reaction outside the cavity involving electron transfer from a reactant potential energy surface with zero vibrational excitations to a product potential energy surface with a single vibrational excitation. The reaction rate is then determined by the competing forward rate from reactant to product, backward rate from product to reactant, and thermalization rate from the product potential energy surfaces with one vibrational excitation to zero. When the thermalization rate is much higher than the backwards rate, the electron cannot transfer back from the product to the reactant. Then, they show that, in the cavity with $N$ disordered molecular vibrations, the dark states become localized on 2-3 molecules. The consequence to the electron transfer rates is that, while the forward rate decreases slightly and the backwards rate is largely unchanged, the thermalization rate increases drastically, as in Fig. \ref{fig:theories}(d) and  overall increasing the reaction rate. Importantly, this theory so far only leads to acceleration of reaction rates, whereas reactions in experiments often slow down in cavities. These opposing trends are not entirely irreconcilable, as knowledge of the exact reaction mechanism is generally necessary to confirm how changing the rate of one elementary step changes the overall reaction rate. Note that this theory is demonstrated through the example of an electron transfer-type reaction that does not correspond to the reaction in the seminal experiment; the authors, however, claim that a similar mechanism could work in adiabatic reactions and plan to publish such theoretical evidence in the near future.

Another flavor of theory that explains only the $N$-dependence has been discussed in both Refs. \citenum{Schutz2020} and \citenum{Sidler2021} by Schutz \textit{et al.} and Sidler \textit{et al.}, respectively, where they show that ensembles of emitters or molecules can induce modifications of optical and chemical properties on the local level. More specifically, in Ref. \citenum{Schutz2020}, they consider a single quantum emitter A within a wavelength of an ensemble B of $N$ emitters all coupled together \textit{via} position- and orientation-dependent dipole-dipole coupling, as in Fig. \ref{fig:theories}(e). They show that, by interacting through the virtual excitations of the ensemble, the effective coupling strength between A and a cavity mode can be increased and the effective linewidth of A can be decreased by modulating the detuning between A and B, enough to put A in the strong-coupling regime for a silicon vacancy defect in diamond. In Ref. \citenum{Sidler2021}, the authors show from first principles that in an ensemble of nitrogen dimers collectively coupled to a cavity mode, the transition dipole moment of an impurity, or a nitrogen dimer with perturbed bond length, increases with the number of molecules in the ensemble. While these theories have not yet been extended to chemical reactions, these recent efforts represent promising steps to an explanation of the $N$-dependence and warrant further study.

\section{Outlook} 
\label{sec:outlook}

So far, we have reviewed the most important experiments, extracting three robust features of vibrational polariton chemistry: the $N$-dependence, resonance, and symmetry effects. We then reviewed several theories that have attempted to explain some of these effects. What open questions are left, and what next steps can we take, as a community, to answer them?

First, a very fundamental question: Is entering the strong coupling regime even a necessary prerequisite for observing cavity-modified ground-state chemical reactivity? In their seminal experiment, Thomas \textit{et al.} observe that increasing Rabi splitting accompanies a decreasing rate constant and conclude that strong light-matter coupling is necessary to observe cavity-modified ground-state chemical reactivity. While there has been some evidence that there is, in fact, a need to enter the strong coupling regime (see, \textit{e.g.}, Ref. \cite{Pang2020}), there has also been evidence to the contrary (see, \textit{e.g.}, Ref. \cite{Vurgaftman2020}). More generally, while a non-zero Rabi splitting, where light-matter coupling is large enough to overcome system losses and decoherence, is a valuable experimental signature of strong light-matter coupling, its presence does not imply that another observed phenomenon relies on it. The Rabi splitting is not only a function of the number of particles $N$, or the concentration of reactant here, but also the vacuum electric field strength of the cavity and the vibrational transition dipole moment. Therefore, one could observe identical Rabi splittings between systems with, say, a larger vacuum electric strength and smaller concentration of reactant versus a smaller vacuum electric field strength and larger concentration of reactant. To more convincingly demonstrate that strong light-matter coupling is a prerequisite of the changes to the reaction rate constant observed, further experiments should explore the effects of changing the linewidths of the cavity modes and vacuum electric strengths to explore the gamut of low-to-high system losses and light-matter coupling strengths. Ideally, determining the effect of changing the vibrational transition dipole moment of the molecule in interest would also be a valuable data point, but doing so is likely more difficult, as it would require changing the reactant molecules themselves and may potentially lead to, for instance, a different reaction mechanism. In parallel, theorists should re-visit already-developed theories to see if similar effects can be seen throughout the parameter space of system loss and light-matter coupling strength, as well as consider theories that do not explicitly depend on the strong coupling regime by including non-zero cavity losses and vibrational linewidths in the foundation of their theories.

Second, experimentalists have observed large changes in activation energies in cavity-modified ground-state chemical reactions that are sometimes indicative of a change in reaction mechanism. Do these changes actually correspond to new reaction pathways, or are these simply effective changes and the original pathways are simply being accelerated or decelerated by the cavity mode? Such studies could be conducted with standard techniques of synthetic chemistry, such as capturing the reaction intermediate to determine whether the reaction has truly transformed from an associative to dissociative nature. Answering each of these questions would be invaluable to theorists, especially in narrowing down which interactions are necessary to include in the system Hamiltonian and whether to include different reaction types and pathways in a chemical dynamics model.

Third, what are the origins of the $N$-dependence, resonance, and symmetry effects? Regarding the $N$-dependence, while there certainly have been jumps in theoretical progress as discussed in Section \ref{sec:theories}, both experimental testing of these theories and extending these theories to the context of chemical reaction dynamics are key. On the theoretical side, it is clear from work thus far that the conditions for collective enhancement of local properties have rather stringent requirements with respect to the detunings and inter-molecular distances and orientations. Future studies should carefully consider the orientation- and geometry-dependence of the surrounding molecules and whether these effects are robust in a liquid environment at room temperature, as in the experiments. On a more general level, it seems that having the molecules oriented in some concerted fashion and not randomly inside the cavity is key to observing several proposed explanations for the $N$-dependence. Are we absolutely convinced that no molecular ordering within the Fabry-Perot cavities is present? Can cavities can be constructed that enhance or stifle molecular ordering to test how modifications to the chemical reactivity scale with molecular ordering?

Regarding the resonance effect, Ref. \citenum{Schafer2019} is so far the only theoretical or computational study that has correctly predicted the frequency of the resonance effect, and therefore, it is the author's opinion that the next major stop on the road toward the theory of vibrational polariton chemistry is better understanding cavity-modified vibrational energy redistribution that evidently leads to the observed ``bond-strengthening" effect. To tease out whether cavity-modified vibrational energy redistribution is responsible for the observed changes in the reaction rates inside cavities, both theorists and experimentalists should first study molecules with fewer numbers of atoms $M$ and, therefore, simpler vibrational structures, as the number of vibrational modes for a nonlinear molecule is $3M-6$. These molecules can be selected to have vibrations of modes that either do or do not involve atoms in the bond of interest, are either close or far in frequency from the vibration involving the bond of interest, and are either nonlinearly coupled or not to the vibration involving the bond of interest. Further studying IVR naturally dovetails with unraveling the symmetry effect because, to inform molecule choice before running costly experiments, vibrational modes and nonlinear couplings can be derived from symmetry analyses, as well as calculated from first principles. Previous experimental studies can also be re-analyzed through this perspective---perhaps, for instance, the resonant cavity frequencies that result in reaction selectivity in Ref. \citenum{Thomas2019} correspond to certain vibrational modes that are coupled to the vibrational mode relevant to the reaction, thus dissipating the vibrational energy. Community members should keep in mind that some vibrational modes are symmetry-forbidden from appearing in IR spectra but may still couple to other vibrational modes.

Looking even further, as the seminal experiment becomes better understood with a robust theory, the field of vibrational polariton chemistry will still have much room to grow. For instance, the ability to either selectively speed up or slow down a given reaction would be useful, and some first steps in this direction could include concentrating vibrational energy into a single mode by cavity engineering or pulse shaping. In addition, different chemical reaction types can be explored by coupling to different vacuum modes, such as vacuum magnetic fields in magnonic cavities \cite{Neuman2020, Wang2021} to influence chemical reactions involving singlet-triplet transitions and free radicals or to chiral cavities for enantioselective chemistry.

Finally, as this work is not meant to be a comprehensive review, which can instead be found in Ref. \citenum{Feist2018, Flick2018, Ribeiro2018, Herrera2020}, but rather a compact perspective of a burgeoning field, some studies will have necessarily been excluded or missed. We appreciate all feedback and are eager to update this work should including any of these excluded or missing papers be constructive toward the arguments made in this work. On a similar note, given the speed at which the field is moving and how much is still unknown, we encourage community members to publicly release as much information as possible about their work, including code for complex simulations and details on both successful and ``failed" experiments on public-facing venues (especially given the difficulty of reproducing some studies \cite{Imperatore2021}), such as ChemRXiv and arXiv, online forums for polariton chemistry, data banks, and research group websites. The more the community knows, the faster we will reach our destination: the theory of vibrational polariton chemistry.

\section*{Acknowledgments}
The authors acknowledge valuable discussions with Tom\'{a}\v{s} Neuman, Johannes Flick, Arkajit Mandal, and Vamsi Varanasi. D.S.W. is an NSF Graduate Research Fellow. S.F.Y. would like to thank the Department of Energy for funding under award DE-SC0020115 and the NSF via the CUA PFC award.


\begin{thebibliography}{39}%
\makeatletter
\providecommand \@ifxundefined [1]{%
 \@ifx{#1\undefined}
}%
\providecommand \@ifnum [1]{%
 \ifnum #1\expandafter \@firstoftwo
 \else \expandafter \@secondoftwo
 \fi
}%
\providecommand \@ifx [1]{%
 \ifx #1\expandafter \@firstoftwo
 \else \expandafter \@secondoftwo
 \fi
}%
\providecommand \natexlab [1]{#1}%
\providecommand \enquote  [1]{``#1''}%
\providecommand \bibnamefont  [1]{#1}%
\providecommand \bibfnamefont [1]{#1}%
\providecommand \citenamefont [1]{#1}%
\providecommand \href@noop [0]{\@secondoftwo}%
\providecommand \href [0]{\begingroup \@sanitize@url \@href}%
\providecommand \@href[1]{\@@startlink{#1}\@@href}%
\providecommand \@@href[1]{\endgroup#1\@@endlink}%
\providecommand \@sanitize@url [0]{\catcode `\\12\catcode `\$12\catcode
  `\&12\catcode `\#12\catcode `\^12\catcode `\_12\catcode `\%12\relax}%
\providecommand \@@startlink[1]{}%
\providecommand \@@endlink[0]{}%
\providecommand \url  [0]{\begingroup\@sanitize@url \@url }%
\providecommand \@url [1]{\endgroup\@href {#1}{\urlprefix }}%
\providecommand \urlprefix  [0]{URL }%
\providecommand \Eprint [0]{\href }%
\providecommand \doibase [0]{https://doi.org/}%
\providecommand \selectlanguage [0]{\@gobble}%
\providecommand \bibinfo  [0]{\@secondoftwo}%
\providecommand \bibfield  [0]{\@secondoftwo}%
\providecommand \translation [1]{[#1]}%
\providecommand \BibitemOpen [0]{}%
\providecommand \bibitemStop [0]{}%
\providecommand \bibitemNoStop [0]{.\EOS\space}%
\providecommand \EOS [0]{\spacefactor3000\relax}%
\providecommand \BibitemShut  [1]{\csname bibitem#1\endcsname}%
\let\auto@bib@innerbib\@empty
\bibitem [{\citenamefont {Frei}\ \emph {et~al.}(1980)\citenamefont {Frei},
  \citenamefont {Fredin},\ and\ \citenamefont {Pimentel}}]{Frei1980}%
  \BibitemOpen
  \bibfield  {author} {\bibinfo {author} {\bibfnamefont {H.}~\bibnamefont
  {Frei}}, \bibinfo {author} {\bibfnamefont {L.}~\bibnamefont {Fredin}},\ and\
  \bibinfo {author} {\bibfnamefont {G.~C.}\ \bibnamefont {Pimentel}},\
  }\bibfield  {title} {\bibinfo {title} {{Vibrational excitation of ozone and
  molecular fluorine reactions in cryogenic matrices}},\ }\href
  {https://doi.org/10.1063/1.440846} {\bibfield  {journal} {\bibinfo  {journal}
  {Journal of Chemical Physics}\ }\textbf {\bibinfo {volume} {74}},\ \bibinfo
  {pages} {397} (\bibinfo {year} {1980})}\BibitemShut {NoStop}%
\bibitem [{\citenamefont {Frei}\ and\ \citenamefont
  {Pimentel}(1983)}]{Frei1983}%
  \BibitemOpen
  \bibfield  {author} {\bibinfo {author} {\bibfnamefont {H.}~\bibnamefont
  {Frei}}\ and\ \bibinfo {author} {\bibfnamefont {G.~C.}\ \bibnamefont
  {Pimentel}},\ }\bibfield  {title} {\bibinfo {title} {{Selective vibrational
  excitation of the ethylene-fluorine reaction in a nitrogen matrix. I}},\
  }\href {https://doi.org/10.1063/1.445825} {\bibfield  {journal} {\bibinfo
  {journal} {Journal of Chemical Physics}\ }\textbf {\bibinfo {volume} {7}},\
  \bibinfo {pages} {3698} (\bibinfo {year} {1983})}\BibitemShut {NoStop}%
\bibitem [{\citenamefont {Frei}(1983)}]{Frei1983a}%
  \BibitemOpen
  \bibfield  {author} {\bibinfo {author} {\bibfnamefont {H.}~\bibnamefont
  {Frei}},\ }\bibfield  {title} {\bibinfo {title} {{Selective vibrational
  excitation of the ethylene-fluorine reaction in a nitrogen matrix. II}},\
  }\href {https://doi.org/10.1063/1.445825} {\bibfield  {journal} {\bibinfo
  {journal} {Journal of Chemical Physics}\ }\textbf {\bibinfo {volume} {79}},\
  \bibinfo {pages} {748} (\bibinfo {year} {1983})}\BibitemShut {NoStop}%
\bibitem [{\citenamefont {Bucksbaum}\ \emph {et~al.}(1990)\citenamefont
  {Bucksbaum}, \citenamefont {Zavriyev}, \citenamefont {Muller},\ and\
  \citenamefont {Schumacher}}]{Bucksbaum1990}%
  \BibitemOpen
  \bibfield  {author} {\bibinfo {author} {\bibfnamefont {P.~H.}\ \bibnamefont
  {Bucksbaum}}, \bibinfo {author} {\bibfnamefont {A.}~\bibnamefont {Zavriyev}},
  \bibinfo {author} {\bibfnamefont {H.~G.}\ \bibnamefont {Muller}},\ and\
  \bibinfo {author} {\bibfnamefont {D.~W.}\ \bibnamefont {Schumacher}},\
  }\bibfield  {title} {\bibinfo {title} {{Softening of the H$_2^+$ Molecular
  Bond in Intense Laser Fields}},\ }\href
  {https://link.aps.org/doi/10.1103/PhysRevLett.64.1931{\%}0Ahttps://link.aps.org/doi/10.1103/PhysRevLett.64.1883}
  {\bibfield  {journal} {\bibinfo  {journal} {Physical Review Letters}\
  }\textbf {\bibinfo {volume} {64}},\ \bibinfo {pages} {1883} (\bibinfo {year}
  {1990})}\BibitemShut {NoStop}%
\bibitem [{\citenamefont {Thomas}\ \emph {et~al.}(2016)\citenamefont {Thomas},
  \citenamefont {George}, \citenamefont {Shalabney}, \citenamefont {Dryzhakov},
  \citenamefont {Varma}, \citenamefont {Moran}, \citenamefont {Chervy},
  \citenamefont {Zhong}, \citenamefont {Devaux}, \citenamefont {Genet},
  \citenamefont {Hutchison},\ and\ \citenamefont {Ebbesen}}]{Thomas2016}%
  \BibitemOpen
  \bibfield  {author} {\bibinfo {author} {\bibfnamefont {A.}~\bibnamefont
  {Thomas}}, \bibinfo {author} {\bibfnamefont {J.}~\bibnamefont {George}},
  \bibinfo {author} {\bibfnamefont {A.}~\bibnamefont {Shalabney}}, \bibinfo
  {author} {\bibfnamefont {M.}~\bibnamefont {Dryzhakov}}, \bibinfo {author}
  {\bibfnamefont {S.~J.}\ \bibnamefont {Varma}}, \bibinfo {author}
  {\bibfnamefont {J.}~\bibnamefont {Moran}}, \bibinfo {author} {\bibfnamefont
  {T.}~\bibnamefont {Chervy}}, \bibinfo {author} {\bibfnamefont
  {X.}~\bibnamefont {Zhong}}, \bibinfo {author} {\bibfnamefont
  {E.}~\bibnamefont {Devaux}}, \bibinfo {author} {\bibfnamefont
  {C.}~\bibnamefont {Genet}}, \bibinfo {author} {\bibfnamefont {J.~A.}\
  \bibnamefont {Hutchison}},\ and\ \bibinfo {author} {\bibfnamefont {T.~W.}\
  \bibnamefont {Ebbesen}},\ }\bibfield  {title} {\bibinfo {title}
  {{Ground-State Chemical Reactivity under Vibrational Coupling to the Vacuum
  Electromagnetic Field}},\ }\href {https://doi.org/10.1002/anie.201605504}
  {\bibfield  {journal} {\bibinfo  {journal} {Angewandte Chemie - International
  Edition}\ }\textbf {\bibinfo {volume} {55}},\ \bibinfo {pages} {11462}
  (\bibinfo {year} {2016})}\BibitemShut {NoStop}%
\bibitem [{\citenamefont {Pang}\ \emph {et~al.}(2020)\citenamefont {Pang},
  \citenamefont {Thomas}, \citenamefont {Nagarajan}, \citenamefont {Vergauwe},
  \citenamefont {Joseph}, \citenamefont {Patrahau}, \citenamefont {Wang},
  \citenamefont {Genet},\ and\ \citenamefont {Ebbesen}}]{Pang2020}%
  \BibitemOpen
  \bibfield  {author} {\bibinfo {author} {\bibfnamefont {Y.}~\bibnamefont
  {Pang}}, \bibinfo {author} {\bibfnamefont {A.}~\bibnamefont {Thomas}},
  \bibinfo {author} {\bibfnamefont {K.}~\bibnamefont {Nagarajan}}, \bibinfo
  {author} {\bibfnamefont {R.~M.~A.}\ \bibnamefont {Vergauwe}}, \bibinfo
  {author} {\bibfnamefont {K.}~\bibnamefont {Joseph}}, \bibinfo {author}
  {\bibfnamefont {B.}~\bibnamefont {Patrahau}}, \bibinfo {author}
  {\bibfnamefont {K.}~\bibnamefont {Wang}}, \bibinfo {author} {\bibfnamefont
  {C.}~\bibnamefont {Genet}},\ and\ \bibinfo {author} {\bibfnamefont {T.~W.}\
  \bibnamefont {Ebbesen}},\ }\bibfield  {title} {\bibinfo {title} {{On the Role
  of Symmetry in Vibrational Strong Coupling: The Case of Charge‐Transfer
  Complexation}},\ }\href {https://doi.org/10.1002/ange.202002527} {\bibfield
  {journal} {\bibinfo  {journal} {Angewandte Chemie}\ }\textbf {\bibinfo
  {volume} {132}},\ \bibinfo {pages} {10522} (\bibinfo {year}
  {2020})}\BibitemShut {NoStop}%
\bibitem [{\citenamefont {Hirai}\ \emph {et~al.}(2020)\citenamefont {Hirai},
  \citenamefont {Takeda}, \citenamefont {Hutchison},\ and\ \citenamefont
  {Uji-i}}]{Hirai2020}%
  \BibitemOpen
  \bibfield  {author} {\bibinfo {author} {\bibfnamefont {K.}~\bibnamefont
  {Hirai}}, \bibinfo {author} {\bibfnamefont {R.}~\bibnamefont {Takeda}},
  \bibinfo {author} {\bibfnamefont {J.~A.}\ \bibnamefont {Hutchison}},\ and\
  \bibinfo {author} {\bibfnamefont {H.}~\bibnamefont {Uji-i}},\ }\bibfield
  {title} {\bibinfo {title} {{Modulation of Prins Cyclization by Vibrational
  Strong Coupling}},\ }\href {https://doi.org/10.1002/anie.201915632}
  {\bibfield  {journal} {\bibinfo  {journal} {Angewandte Chemie - International
  Edition}\ }\textbf {\bibinfo {volume} {59}},\ \bibinfo {pages} {5332}
  (\bibinfo {year} {2020})}\BibitemShut {NoStop}%
\bibitem [{\citenamefont {Sau}\ \emph {et~al.}(2021)\citenamefont {Sau},
  \citenamefont {Nagarajan}, \citenamefont {Patrahau}, \citenamefont
  {Lethuillier‐Karl}, \citenamefont {Vergauwe}, \citenamefont {Thomas},
  \citenamefont {Moran}, \citenamefont {Genet},\ and\ \citenamefont
  {Ebbesen}}]{Sau2021}%
  \BibitemOpen
  \bibfield  {author} {\bibinfo {author} {\bibfnamefont {A.}~\bibnamefont
  {Sau}}, \bibinfo {author} {\bibfnamefont {K.}~\bibnamefont {Nagarajan}},
  \bibinfo {author} {\bibfnamefont {B.}~\bibnamefont {Patrahau}}, \bibinfo
  {author} {\bibfnamefont {L.}~\bibnamefont {Lethuillier‐Karl}}, \bibinfo
  {author} {\bibfnamefont {R.~M.~A.}\ \bibnamefont {Vergauwe}}, \bibinfo
  {author} {\bibfnamefont {A.}~\bibnamefont {Thomas}}, \bibinfo {author}
  {\bibfnamefont {J.}~\bibnamefont {Moran}}, \bibinfo {author} {\bibfnamefont
  {C.}~\bibnamefont {Genet}},\ and\ \bibinfo {author} {\bibfnamefont {T.~W.}\
  \bibnamefont {Ebbesen}},\ }\bibfield  {title} {\bibinfo {title} {{Modifying
  Woodward–Hoffmann Stereoselectivity Under Vibrational Strong Coupling}},\
  }\href {https://doi.org/10.1002/ange.202013465} {\bibfield  {journal}
  {\bibinfo  {journal} {Angewandte Chemie}\ }\textbf {\bibinfo {volume}
  {133}},\ \bibinfo {pages} {5776} (\bibinfo {year} {2021})}\BibitemShut
  {NoStop}%
\bibitem [{\citenamefont {Galego}\ \emph {et~al.}(2019)\citenamefont {Galego},
  \citenamefont {Climent}, \citenamefont {Garcia-Vidal},\ and\ \citenamefont
  {Feist}}]{Galego2019}%
  \BibitemOpen
  \bibfield  {author} {\bibinfo {author} {\bibfnamefont {J.}~\bibnamefont
  {Galego}}, \bibinfo {author} {\bibfnamefont {C.}~\bibnamefont {Climent}},
  \bibinfo {author} {\bibfnamefont {F.~J.}\ \bibnamefont {Garcia-Vidal}},\ and\
  \bibinfo {author} {\bibfnamefont {J.}~\bibnamefont {Feist}},\ }\bibfield
  {title} {\bibinfo {title} {{Cavity Casimir-Polder Forces and Their Effects in
  Ground-State Chemical Reactivity}},\ }\href
  {https://doi.org/10.1103/PhysRevX.9.021057} {\bibfield  {journal} {\bibinfo
  {journal} {Physical Review X}\ }\textbf {\bibinfo {volume} {9}},\ \bibinfo
  {pages} {021057} (\bibinfo {year} {2019})}\BibitemShut {NoStop}%
\bibitem [{\citenamefont {Campos-Gonzalez-Angulo}\ \emph
  {et~al.}(2019)\citenamefont {Campos-Gonzalez-Angulo}, \citenamefont
  {Ribeiro},\ and\ \citenamefont {Yuen-Zhou}}]{Campos-Gonzalez-Angulo2019}%
  \BibitemOpen
  \bibfield  {author} {\bibinfo {author} {\bibfnamefont {J.~A.}\ \bibnamefont
  {Campos-Gonzalez-Angulo}}, \bibinfo {author} {\bibfnamefont {R.~F.}\
  \bibnamefont {Ribeiro}},\ and\ \bibinfo {author} {\bibfnamefont
  {J.}~\bibnamefont {Yuen-Zhou}},\ }\bibfield  {title} {\bibinfo {title}
  {{Resonant catalysis of thermally activated chemical reactions with
  vibrational polaritons}},\ }\href
  {https://doi.org/10.1038/s41467-019-12636-1} {\bibfield  {journal} {\bibinfo
  {journal} {Nature Communications}\ }\textbf {\bibinfo {volume} {10}},\
  \bibinfo {pages} {1} (\bibinfo {year} {2019})}\BibitemShut {NoStop}%
\bibitem [{\citenamefont {Li}\ \emph {et~al.}(2021{\natexlab{a}})\citenamefont
  {Li}, \citenamefont {Mandal},\ and\ \citenamefont {Huo}}]{Li2021a}%
  \BibitemOpen
  \bibfield  {author} {\bibinfo {author} {\bibfnamefont {X.}~\bibnamefont
  {Li}}, \bibinfo {author} {\bibfnamefont {A.}~\bibnamefont {Mandal}},\ and\
  \bibinfo {author} {\bibfnamefont {P.}~\bibnamefont {Huo}},\ }\bibfield
  {title} {\bibinfo {title} {{Cavity frequency-dependent theory for vibrational
  polariton chemistry}},\ }\href {https://doi.org/10.1038/s41467-021-21610-9}
  {\bibfield  {journal} {\bibinfo  {journal} {Nature Communications}\ }\textbf
  {\bibinfo {volume} {12}},\ \bibinfo {pages} {1} (\bibinfo {year}
  {2021}{\natexlab{a}})}\BibitemShut {NoStop}%
\bibitem [{\citenamefont {Du}\ and\ \citenamefont {Yuen-Zhou}(2021)}]{Du2021}%
  \BibitemOpen
  \bibfield  {author} {\bibinfo {author} {\bibfnamefont {M.}~\bibnamefont
  {Du}}\ and\ \bibinfo {author} {\bibfnamefont {J.}~\bibnamefont {Yuen-Zhou}},\
  }\bibfield  {title} {\bibinfo {title} {{Can Dark States Explain
  Vibropolaritonic Chemistry?}},\ }\href {http://arxiv.org/abs/2104.07214}
  {\bibfield  {journal} {\bibinfo  {journal} {arXiv:2104.07214}\ } (\bibinfo
  {year} {2021})}\BibitemShut {NoStop}%
\bibitem [{\citenamefont {Sch{\"{u}}tz}\ \emph {et~al.}(2020)\citenamefont
  {Sch{\"{u}}tz}, \citenamefont {Schachenmayer}, \citenamefont
  {Hagenm{\"{u}}ller}, \citenamefont {Brennen}, \citenamefont {Volz},
  \citenamefont {Sandoghdar}, \citenamefont {Ebbesen}, \citenamefont {Genes},\
  and\ \citenamefont {Pupillo}}]{Schutz2020}%
  \BibitemOpen
  \bibfield  {author} {\bibinfo {author} {\bibfnamefont {S.}~\bibnamefont
  {Sch{\"{u}}tz}}, \bibinfo {author} {\bibfnamefont {J.}~\bibnamefont
  {Schachenmayer}}, \bibinfo {author} {\bibfnamefont {D.}~\bibnamefont
  {Hagenm{\"{u}}ller}}, \bibinfo {author} {\bibfnamefont {G.~K.}\ \bibnamefont
  {Brennen}}, \bibinfo {author} {\bibfnamefont {T.}~\bibnamefont {Volz}},
  \bibinfo {author} {\bibfnamefont {V.}~\bibnamefont {Sandoghdar}}, \bibinfo
  {author} {\bibfnamefont {T.~W.}\ \bibnamefont {Ebbesen}}, \bibinfo {author}
  {\bibfnamefont {C.}~\bibnamefont {Genes}},\ and\ \bibinfo {author}
  {\bibfnamefont {G.}~\bibnamefont {Pupillo}},\ }\bibfield  {title} {\bibinfo
  {title} {{Ensemble-Induced Strong Light-Matter Coupling of a Single Quantum
  Emitter}},\ }\href {https://doi.org/10.1103/PhysRevLett.124.113602}
  {\bibfield  {journal} {\bibinfo  {journal} {Physical Review Letters}\
  }\textbf {\bibinfo {volume} {124}},\ \bibinfo {pages} {113602} (\bibinfo
  {year} {2020})}\BibitemShut {NoStop}%
\bibitem [{\citenamefont {Thomas}\ \emph {et~al.}(2020)\citenamefont {Thomas},
  \citenamefont {Jayachandran}, \citenamefont {Lethuillier-Karl}, \citenamefont
  {Vergauwe}, \citenamefont {Nagarajan}, \citenamefont {Devaux}, \citenamefont
  {Genet}, \citenamefont {Moran},\ and\ \citenamefont {Ebbesen}}]{Thomas2020}%
  \BibitemOpen
  \bibfield  {author} {\bibinfo {author} {\bibfnamefont {A.}~\bibnamefont
  {Thomas}}, \bibinfo {author} {\bibfnamefont {A.}~\bibnamefont
  {Jayachandran}}, \bibinfo {author} {\bibfnamefont {L.}~\bibnamefont
  {Lethuillier-Karl}}, \bibinfo {author} {\bibfnamefont {R.~M.}\ \bibnamefont
  {Vergauwe}}, \bibinfo {author} {\bibfnamefont {K.}~\bibnamefont {Nagarajan}},
  \bibinfo {author} {\bibfnamefont {E.}~\bibnamefont {Devaux}}, \bibinfo
  {author} {\bibfnamefont {C.}~\bibnamefont {Genet}}, \bibinfo {author}
  {\bibfnamefont {J.}~\bibnamefont {Moran}},\ and\ \bibinfo {author}
  {\bibfnamefont {T.~W.}\ \bibnamefont {Ebbesen}},\ }\bibfield  {title}
  {\bibinfo {title} {{Ground state chemistry under vibrational strong coupling:
  Dependence of thermodynamic parameters on the Rabi splitting energy}},\
  }\href {https://doi.org/10.1515/nanoph-2019-0340} {\bibfield  {journal}
  {\bibinfo  {journal} {Nanophotonics}\ }\textbf {\bibinfo {volume} {9}},\
  \bibinfo {pages} {249} (\bibinfo {year} {2020})}\BibitemShut {NoStop}%
\bibitem [{\citenamefont {Miller}(1974)}]{Miller1974}%
  \BibitemOpen
  \bibfield  {author} {\bibinfo {author} {\bibfnamefont {W.~H.}\ \bibnamefont
  {Miller}},\ }\bibfield  {title} {\bibinfo {title} {{Quantum mechanical
  transition state theory and a new semiclassical model for reaction rate
  constants}},\ }\href {https://doi.org/10.1063/1.1682181} {\bibfield
  {journal} {\bibinfo  {journal} {Journal of Chemical Physics}\ }\textbf
  {\bibinfo {volume} {61}},\ \bibinfo {pages} {1823} (\bibinfo {year}
  {1974})}\BibitemShut {NoStop}%
\bibitem [{\citenamefont {Miller}\ \emph {et~al.}(1983)\citenamefont {Miller},
  \citenamefont {Schwartz},\ and\ \citenamefont {Tromp}}]{Miller1983}%
  \BibitemOpen
  \bibfield  {author} {\bibinfo {author} {\bibfnamefont {W.~H.}\ \bibnamefont
  {Miller}}, \bibinfo {author} {\bibfnamefont {S.~D.}\ \bibnamefont
  {Schwartz}},\ and\ \bibinfo {author} {\bibfnamefont {J.~W.}\ \bibnamefont
  {Tromp}},\ }\bibfield  {title} {\bibinfo {title} {{Quantum mechanical rate
  constants for bimolecular reactions}},\ }\href
  {https://doi.org/10.1063/1.445581} {\bibfield  {journal} {\bibinfo  {journal}
  {Journal of Chemical Physics}\ }\textbf {\bibinfo {volume} {79}},\ \bibinfo
  {pages} {4889} (\bibinfo {year} {1983})}\BibitemShut {NoStop}%
\bibitem [{Man()}]{Manolopoulous2008}%
  \BibitemOpen
  \href@noop {} {\bibinfo {title} {Chemical reaction dynamics}},\ \bibinfo
  {howpublished} {\url{http://manolopoulos.chem.ox.ac.uk/downloads/2008.pdf}},\
  \bibinfo {note} {accessed: 2021-05-03}\BibitemShut {NoStop}%
\bibitem [{\citenamefont {Li}\ \emph {et~al.}(2020{\natexlab{a}})\citenamefont
  {Li}, \citenamefont {Subotnik},\ and\ \citenamefont {Nitzan}}]{Li2020}%
  \BibitemOpen
  \bibfield  {author} {\bibinfo {author} {\bibfnamefont {T.~E.}\ \bibnamefont
  {Li}}, \bibinfo {author} {\bibfnamefont {J.~E.}\ \bibnamefont {Subotnik}},\
  and\ \bibinfo {author} {\bibfnamefont {A.}~\bibnamefont {Nitzan}},\
  }\bibfield  {title} {\bibinfo {title} {{Cavity molecular dynamics simulations
  of liquid water under vibrational ultrastrong coupling}},\ }\href
  {https://doi.org/10.1073/pnas.2009272117} {\bibfield  {journal} {\bibinfo
  {journal} {Proceedings of the National Academy of Sciences of the United
  States of America}\ }\textbf {\bibinfo {volume} {117}},\ \bibinfo {pages}
  {18324} (\bibinfo {year} {2020}{\natexlab{a}})},\ \Eprint
  {https://arxiv.org/abs/2004.04888} {arXiv:2004.04888} \BibitemShut {NoStop}%
\bibitem [{\citenamefont {Li}\ \emph {et~al.}(2020{\natexlab{b}})\citenamefont
  {Li}, \citenamefont {Nitzan},\ and\ \citenamefont {Subotnik}}]{Li2020a}%
  \BibitemOpen
  \bibfield  {author} {\bibinfo {author} {\bibfnamefont {T.~E.}\ \bibnamefont
  {Li}}, \bibinfo {author} {\bibfnamefont {A.}~\bibnamefont {Nitzan}},\ and\
  \bibinfo {author} {\bibfnamefont {J.~E.}\ \bibnamefont {Subotnik}},\
  }\bibfield  {title} {\bibinfo {title} {{On the origin of ground-state
  vacuum-field catalysis: Equilibrium consideration}},\ }\href
  {https://doi.org/10.1063/5.0006472} {\bibfield  {journal} {\bibinfo
  {journal} {Journal of Chemical Physics}\ }\textbf {\bibinfo {volume} {152}},\
  \bibinfo {pages} {234107} (\bibinfo {year} {2020}{\natexlab{b}})}\BibitemShut
  {NoStop}%
\bibitem [{\citenamefont {Vurgaftman}\ \emph {et~al.}(2020)\citenamefont
  {Vurgaftman}, \citenamefont {Simpkins}, \citenamefont {Dunkelberger},\ and\
  \citenamefont {Owrutsky}}]{Vurgaftman2020}%
  \BibitemOpen
  \bibfield  {author} {\bibinfo {author} {\bibfnamefont {I.}~\bibnamefont
  {Vurgaftman}}, \bibinfo {author} {\bibfnamefont {B.~S.}\ \bibnamefont
  {Simpkins}}, \bibinfo {author} {\bibfnamefont {A.~D.}\ \bibnamefont
  {Dunkelberger}},\ and\ \bibinfo {author} {\bibfnamefont {J.~C.}\ \bibnamefont
  {Owrutsky}},\ }\bibfield  {title} {\bibinfo {title} {{Negligible Effect of
  Vibrational Polaritons on Chemical Reaction Rates via the Density of States
  Pathway}},\ }\href {https://doi.org/10.1021/acs.jpclett.0c00841} {\bibfield
  {journal} {\bibinfo  {journal} {The Journal of Physical Chemistry Letters}\
  }\textbf {\bibinfo {volume} {11}},\ \bibinfo {pages} {3557} (\bibinfo {year}
  {2020})}\BibitemShut {NoStop}%
\bibitem [{\citenamefont {Li}\ \emph {et~al.}(2021{\natexlab{b}})\citenamefont
  {Li}, \citenamefont {Mandal},\ and\ \citenamefont {Huo}}]{Li2021}%
  \BibitemOpen
  \bibfield  {author} {\bibinfo {author} {\bibfnamefont {X.}~\bibnamefont
  {Li}}, \bibinfo {author} {\bibfnamefont {A.}~\bibnamefont {Mandal}},\ and\
  \bibinfo {author} {\bibfnamefont {P.}~\bibnamefont {Huo}},\ }\href
  {https://doi.org/10.26434/chemrxiv.14394815.v1} {\bibinfo {title}
  {Mode-selective chemistry through polaritonic vibrational strong coupling}}
  (\bibinfo {year} {2021}{\natexlab{b}})\BibitemShut {NoStop}%
\bibitem [{\citenamefont {Grote}\ and\ \citenamefont
  {Hynes}(1980)}]{Grote1980}%
  \BibitemOpen
  \bibfield  {author} {\bibinfo {author} {\bibfnamefont {R.~F.}\ \bibnamefont
  {Grote}}\ and\ \bibinfo {author} {\bibfnamefont {J.~T.}\ \bibnamefont
  {Hynes}},\ }\bibfield  {title} {\bibinfo {title} {{The stable states picture
  of chemical reactions. II. Rate constants for condensed and gas phase
  reaction models}},\ }\href {https://doi.org/10.1063/1.440485} {\bibfield
  {journal} {\bibinfo  {journal} {Journal of Chemical Physics}\ }\textbf
  {\bibinfo {volume} {73}},\ \bibinfo {pages} {2715} (\bibinfo {year}
  {1980})}\BibitemShut {NoStop}%
\bibitem [{\citenamefont {H{\"{a}}nggi}\ \emph {et~al.}(1990)\citenamefont
  {H{\"{a}}nggi}, \citenamefont {Talkner},\ and\ \citenamefont
  {Borkovec}}]{Hanggi1990}%
  \BibitemOpen
  \bibfield  {author} {\bibinfo {author} {\bibfnamefont {P.}~\bibnamefont
  {H{\"{a}}nggi}}, \bibinfo {author} {\bibfnamefont {P.}~\bibnamefont
  {Talkner}},\ and\ \bibinfo {author} {\bibfnamefont {M.}~\bibnamefont
  {Borkovec}},\ }\bibfield  {title} {\bibinfo {title} {{Reaction-rate theory:
  Fifty years after Kramers}},\ }\href
  {https://doi.org/10.1103/RevModPhys.62.251} {\bibfield  {journal} {\bibinfo
  {journal} {Reviews of Modern Physics}\ }\textbf {\bibinfo {volume} {62}},\
  \bibinfo {pages} {251} (\bibinfo {year} {1990})}\BibitemShut {NoStop}%
\bibitem [{\citenamefont {Sch{\"{a}}fer}\ \emph {et~al.}(2021)\citenamefont
  {Sch{\"{a}}fer}, \citenamefont {Flick}, \citenamefont {Ronca}, \citenamefont
  {Narang},\ and\ \citenamefont {Rubio}}]{Schafer2021}%
  \BibitemOpen
  \bibfield  {author} {\bibinfo {author} {\bibfnamefont {C.}~\bibnamefont
  {Sch{\"{a}}fer}}, \bibinfo {author} {\bibfnamefont {J.}~\bibnamefont
  {Flick}}, \bibinfo {author} {\bibfnamefont {E.}~\bibnamefont {Ronca}},
  \bibinfo {author} {\bibfnamefont {P.}~\bibnamefont {Narang}},\ and\ \bibinfo
  {author} {\bibfnamefont {A.}~\bibnamefont {Rubio}},\ }\bibfield  {title}
  {\bibinfo {title} {{Shining Light on the Microscopic Resonant Mechanism
  Responsible for Cavity-Mediated Chemical Reactivity}},\ }\href
  {http://arxiv.org/abs/2104.12429} {\bibfield  {journal} {\bibinfo  {journal}
  {arXiv:2104.12429}\ } (\bibinfo {year} {2021})}\BibitemShut {NoStop}%
\bibitem [{\citenamefont {Tokatly}(2013)}]{tokatly2013}%
  \BibitemOpen
  \bibfield  {author} {\bibinfo {author} {\bibfnamefont {I.~V.}\ \bibnamefont
  {Tokatly}},\ }\bibfield  {title} {\bibinfo {title} {Time-dependent density
  functional theory for many-electron systems interacting with cavity
  photons},\ }\href {https://doi.org/10.1103/PhysRevLett.110.233001} {\bibfield
   {journal} {\bibinfo  {journal} {Phys. Rev. Lett.}\ }\textbf {\bibinfo
  {volume} {110}},\ \bibinfo {pages} {233001} (\bibinfo {year}
  {2013})}\BibitemShut {NoStop}%
\bibitem [{\citenamefont {Ruggenthaler}\ \emph {et~al.}(2014)\citenamefont
  {Ruggenthaler}, \citenamefont {Flick}, \citenamefont {Pellegrini},
  \citenamefont {Appel}, \citenamefont {Tokatly},\ and\ \citenamefont
  {Rubio}}]{ruggenthaler2014}%
  \BibitemOpen
  \bibfield  {author} {\bibinfo {author} {\bibfnamefont {M.}~\bibnamefont
  {Ruggenthaler}}, \bibinfo {author} {\bibfnamefont {J.}~\bibnamefont {Flick}},
  \bibinfo {author} {\bibfnamefont {C.}~\bibnamefont {Pellegrini}}, \bibinfo
  {author} {\bibfnamefont {H.}~\bibnamefont {Appel}}, \bibinfo {author}
  {\bibfnamefont {I.~V.}\ \bibnamefont {Tokatly}},\ and\ \bibinfo {author}
  {\bibfnamefont {A.}~\bibnamefont {Rubio}},\ }\bibfield  {title} {\bibinfo
  {title} {Quantum-electrodynamical density-functional theory: Bridging quantum
  optics and electronic-structure theory},\ }\href
  {https://doi.org/10.1103/PhysRevA.90.012508} {\bibfield  {journal} {\bibinfo
  {journal} {Phys. Rev. A}\ }\textbf {\bibinfo {volume} {90}},\ \bibinfo
  {pages} {012508} (\bibinfo {year} {2014})}\BibitemShut {NoStop}%
\bibitem [{\citenamefont {Flick}\ \emph {et~al.}(2019)\citenamefont {Flick},
  \citenamefont {Welakuh}, \citenamefont {Ruggenthaler}, \citenamefont
  {Appel},\ and\ \citenamefont {Rubio}}]{flick2019lmrnqe}%
  \BibitemOpen
  \bibfield  {author} {\bibinfo {author} {\bibfnamefont {J.}~\bibnamefont
  {Flick}}, \bibinfo {author} {\bibfnamefont {D.~M.}\ \bibnamefont {Welakuh}},
  \bibinfo {author} {\bibfnamefont {M.}~\bibnamefont {Ruggenthaler}}, \bibinfo
  {author} {\bibfnamefont {H.}~\bibnamefont {Appel}},\ and\ \bibinfo {author}
  {\bibfnamefont {A.}~\bibnamefont {Rubio}},\ }\bibfield  {title} {\bibinfo
  {title} {Light–matter response in nonrelativistic quantum
  electrodynamics},\ }\href {https://doi.org/10.1021/acsphotonics.9b00768}
  {\bibfield  {journal} {\bibinfo  {journal} {ACS Photonics}\ }\textbf
  {\bibinfo {volume} {6}},\ \bibinfo {pages} {2757} (\bibinfo {year}
  {2019})}\BibitemShut {NoStop}%
\bibitem [{\citenamefont {Flick}\ and\ \citenamefont
  {Narang}(2020)}]{flickexcited}%
  \BibitemOpen
  \bibfield  {author} {\bibinfo {author} {\bibfnamefont {J.}~\bibnamefont
  {Flick}}\ and\ \bibinfo {author} {\bibfnamefont {P.}~\bibnamefont {Narang}},\
  }\bibfield  {title} {\bibinfo {title} {ab initio polaritonic potential-energy
  surfaces for excited-state nanophotonics and polaritonic chemistry},\ }\href
  {https://aip.scitation.org/doi/10.1063/5.0021033} {\bibfield  {journal}
  {\bibinfo  {journal} {J. Chem. Phys.}\ }\textbf {\bibinfo {volume} {153}},\
  \bibinfo {pages} {094116} (\bibinfo {year} {2020})}\BibitemShut {NoStop}%
\bibitem [{\citenamefont {Wang}\ \emph
  {et~al.}(2021{\natexlab{a}})\citenamefont {Wang}, \citenamefont {Neuman},
  \citenamefont {Flick},\ and\ \citenamefont {Narang}}]{Wang2020LossQEDFT}%
  \BibitemOpen
  \bibfield  {author} {\bibinfo {author} {\bibfnamefont {D.~S.}\ \bibnamefont
  {Wang}}, \bibinfo {author} {\bibfnamefont {T.}~\bibnamefont {Neuman}},
  \bibinfo {author} {\bibfnamefont {J.}~\bibnamefont {Flick}},\ and\ \bibinfo
  {author} {\bibfnamefont {P.}~\bibnamefont {Narang}},\ }\bibfield  {title}
  {\bibinfo {title} {{Light–matter interaction of a molecule in a dissipative
  cavity from first principles}},\ }\href
  {https://aip.scitation.org/doi/full/10.1063/5.0036283} {\bibfield  {journal}
  {\bibinfo  {journal} {J. Chem. Phys.}\ }\textbf {\bibinfo {volume} {154}},\
  \bibinfo {pages} {104109} (\bibinfo {year} {2021}{\natexlab{a}})}\BibitemShut
  {NoStop}%
\bibitem [{\citenamefont {Sidler}\ \emph {et~al.}(2021)\citenamefont {Sidler},
  \citenamefont {Sch{\"{a}}fer}, \citenamefont {Ruggenthaler},\ and\
  \citenamefont {Rubio}}]{Sidler2021}%
  \BibitemOpen
  \bibfield  {author} {\bibinfo {author} {\bibfnamefont {D.}~\bibnamefont
  {Sidler}}, \bibinfo {author} {\bibfnamefont {C.}~\bibnamefont
  {Sch{\"{a}}fer}}, \bibinfo {author} {\bibfnamefont {M.}~\bibnamefont
  {Ruggenthaler}},\ and\ \bibinfo {author} {\bibfnamefont {A.}~\bibnamefont
  {Rubio}},\ }\bibfield  {title} {\bibinfo {title} {{Polaritonic Chemistry:
  Collective Strong Coupling Implies Strong Local Modification of Chemical
  Properties}},\ }\href {https://doi.org/10.1021/acs.jpclett.0c03436}
  {\bibfield  {journal} {\bibinfo  {journal} {Journal of Physical Chemistry
  Letters}\ }\textbf {\bibinfo {volume} {12}},\ \bibinfo {pages} {508}
  (\bibinfo {year} {2021})},\ \Eprint {https://arxiv.org/abs/2011.03284}
  {arXiv:2011.03284} \BibitemShut {NoStop}%
\bibitem [{\citenamefont {Sch{\"{a}}fer}\ \emph {et~al.}(2019)\citenamefont
  {Sch{\"{a}}fer}, \citenamefont {Ruggenthaler}, \citenamefont {Appel},\ and\
  \citenamefont {Rubio}}]{Schafer2019}%
  \BibitemOpen
  \bibfield  {author} {\bibinfo {author} {\bibfnamefont {C.}~\bibnamefont
  {Sch{\"{a}}fer}}, \bibinfo {author} {\bibfnamefont {M.}~\bibnamefont
  {Ruggenthaler}}, \bibinfo {author} {\bibfnamefont {H.}~\bibnamefont
  {Appel}},\ and\ \bibinfo {author} {\bibfnamefont {A.}~\bibnamefont {Rubio}},\
  }\bibfield  {title} {\bibinfo {title} {{Modification of excitation and charge
  transfer in cavity quantum-electrodynamical chemistry}},\ }\href
  {https://doi.org/10.1073/pnas.1814178116} {\bibfield  {journal} {\bibinfo
  {journal} {Proceedings of the National Academy of Sciences of the United
  States of America}\ }\textbf {\bibinfo {volume} {116}},\ \bibinfo {pages}
  {4883} (\bibinfo {year} {2019})}\BibitemShut {NoStop}%
\bibitem [{\citenamefont {Thomas}\ \emph {et~al.}(2019)\citenamefont {Thomas},
  \citenamefont {Lethuillier-Karl}, \citenamefont {Nagarajan}, \citenamefont
  {Vergauwe}, \citenamefont {George}, \citenamefont {Chervy}, \citenamefont
  {Shalabney}, \citenamefont {Devaux}, \citenamefont {Genet}, \citenamefont
  {Moran},\ and\ \citenamefont {Ebbesen}}]{Thomas2019}%
  \BibitemOpen
  \bibfield  {author} {\bibinfo {author} {\bibfnamefont {A.}~\bibnamefont
  {Thomas}}, \bibinfo {author} {\bibfnamefont {L.}~\bibnamefont
  {Lethuillier-Karl}}, \bibinfo {author} {\bibfnamefont {K.}~\bibnamefont
  {Nagarajan}}, \bibinfo {author} {\bibfnamefont {R.~M.}\ \bibnamefont
  {Vergauwe}}, \bibinfo {author} {\bibfnamefont {J.}~\bibnamefont {George}},
  \bibinfo {author} {\bibfnamefont {T.}~\bibnamefont {Chervy}}, \bibinfo
  {author} {\bibfnamefont {A.}~\bibnamefont {Shalabney}}, \bibinfo {author}
  {\bibfnamefont {E.}~\bibnamefont {Devaux}}, \bibinfo {author} {\bibfnamefont
  {C.}~\bibnamefont {Genet}}, \bibinfo {author} {\bibfnamefont
  {J.}~\bibnamefont {Moran}},\ and\ \bibinfo {author} {\bibfnamefont {T.~W.}\
  \bibnamefont {Ebbesen}},\ }\bibfield  {title} {\bibinfo {title} {{Tilting a
  ground-state reactivity landscape by vibrational strong coupling}},\ }\href
  {https://doi.org/10.1126/science.aau7742} {\bibfield  {journal} {\bibinfo
  {journal} {Science}\ }\textbf {\bibinfo {volume} {363}},\ \bibinfo {pages}
  {615} (\bibinfo {year} {2019})}\BibitemShut {NoStop}%
\bibitem [{\citenamefont {Neuman}\ \emph {et~al.}(2020)\citenamefont {Neuman},
  \citenamefont {Wang},\ and\ \citenamefont {Narang}}]{Neuman2020}%
  \BibitemOpen
  \bibfield  {author} {\bibinfo {author} {\bibfnamefont {T.}~\bibnamefont
  {Neuman}}, \bibinfo {author} {\bibfnamefont {D.~S.}\ \bibnamefont {Wang}},\
  and\ \bibinfo {author} {\bibfnamefont {P.}~\bibnamefont {Narang}},\
  }\bibfield  {title} {\bibinfo {title} {{Nanomagnonic Cavities for Strong
  Spin-Magnon Coupling and Magnon-Mediated Spin-Spin Interactions}},\ }\href
  {https://doi.org/10.1103/PhysRevLett.125.247702} {\bibfield  {journal}
  {\bibinfo  {journal} {Physical Review Letters}\ }\textbf {\bibinfo {volume}
  {125}},\ \bibinfo {pages} {247702} (\bibinfo {year} {2020})}\BibitemShut
  {NoStop}%
\bibitem [{\citenamefont {Wang}\ \emph
  {et~al.}(2021{\natexlab{b}})\citenamefont {Wang}, \citenamefont {Neuman},\
  and\ \citenamefont {Narang}}]{Wang2021}%
  \BibitemOpen
  \bibfield  {author} {\bibinfo {author} {\bibfnamefont {D.~S.}\ \bibnamefont
  {Wang}}, \bibinfo {author} {\bibfnamefont {T.}~\bibnamefont {Neuman}},\ and\
  \bibinfo {author} {\bibfnamefont {P.}~\bibnamefont {Narang}},\ }\bibfield
  {title} {\bibinfo {title} {{Spin Emitters beyond the Point Dipole
  Approximation in Nanomagnonic Cavities}},\ }\href
  {https://doi.org/10.1021/acs.jpcc.0c11536} {\bibfield  {journal} {\bibinfo
  {journal} {Journal of Physical Chemistry C}\ }\textbf {\bibinfo {volume}
  {125}},\ \bibinfo {pages} {6222} (\bibinfo {year}
  {2021}{\natexlab{b}})}\BibitemShut {NoStop}%
\bibitem [{\citenamefont {Feist}\ \emph {et~al.}(2018)\citenamefont {Feist},
  \citenamefont {Galego},\ and\ \citenamefont {Garcia-Vidal}}]{Feist2018}%
  \BibitemOpen
  \bibfield  {author} {\bibinfo {author} {\bibfnamefont {J.}~\bibnamefont
  {Feist}}, \bibinfo {author} {\bibfnamefont {J.}~\bibnamefont {Galego}},\ and\
  \bibinfo {author} {\bibfnamefont {F.~J.}\ \bibnamefont {Garcia-Vidal}},\
  }\bibfield  {title} {\bibinfo {title} {{Polaritonic Chemistry with Organic
  Molecules}},\ }\href {https://doi.org/10.1021/acsphotonics.7b00680}
  {\bibfield  {journal} {\bibinfo  {journal} {ACS Photonics}\ }\textbf
  {\bibinfo {volume} {5}},\ \bibinfo {pages} {205} (\bibinfo {year}
  {2018})}\BibitemShut {NoStop}%
\bibitem [{\citenamefont {Flick}\ \emph {et~al.}(2018)\citenamefont {Flick},
  \citenamefont {Rivera},\ and\ \citenamefont {Narang}}]{Flick2018}%
  \BibitemOpen
  \bibfield  {author} {\bibinfo {author} {\bibfnamefont {J.}~\bibnamefont
  {Flick}}, \bibinfo {author} {\bibfnamefont {N.}~\bibnamefont {Rivera}},\ and\
  \bibinfo {author} {\bibfnamefont {P.}~\bibnamefont {Narang}},\ }\bibfield
  {title} {\bibinfo {title} {{Strong light-matter coupling in quantum chemistry
  and quantum photonics}},\ }\href {https://doi.org/10.1515/nanoph-2018-0067}
  {\bibfield  {journal} {\bibinfo  {journal} {Nanophotonics}\ }\textbf
  {\bibinfo {volume} {7}},\ \bibinfo {pages} {1479} (\bibinfo {year}
  {2018})}\BibitemShut {NoStop}%
\bibitem [{\citenamefont {Ribeiro}\ \emph {et~al.}(2018)\citenamefont
  {Ribeiro}, \citenamefont {Mart{\'{i}}nez-Mart{\'{i}}nez}, \citenamefont {Du},
  \citenamefont {Campos-Gonzalez-Angulo},\ and\ \citenamefont
  {Yuen-Zhou}}]{Ribeiro2018}%
  \BibitemOpen
  \bibfield  {author} {\bibinfo {author} {\bibfnamefont {R.~F.}\ \bibnamefont
  {Ribeiro}}, \bibinfo {author} {\bibfnamefont {L.~A.}\ \bibnamefont
  {Mart{\'{i}}nez-Mart{\'{i}}nez}}, \bibinfo {author} {\bibfnamefont
  {M.}~\bibnamefont {Du}}, \bibinfo {author} {\bibfnamefont {J.}~\bibnamefont
  {Campos-Gonzalez-Angulo}},\ and\ \bibinfo {author} {\bibfnamefont
  {J.}~\bibnamefont {Yuen-Zhou}},\ }\bibfield  {title} {\bibinfo {title}
  {{Polariton chemistry: controlling molecular dynamics with optical
  cavities}},\ }\href {https://doi.org/10.1039/c8sc01043a} {\bibfield
  {journal} {\bibinfo  {journal} {Chemical Science}\ }\textbf {\bibinfo
  {volume} {9}},\ \bibinfo {pages} {6325} (\bibinfo {year} {2018})}\BibitemShut
  {NoStop}%
\bibitem [{\citenamefont {Herrera}\ and\ \citenamefont
  {Owrutsky}(2020)}]{Herrera2020}%
  \BibitemOpen
  \bibfield  {author} {\bibinfo {author} {\bibfnamefont {F.}~\bibnamefont
  {Herrera}}\ and\ \bibinfo {author} {\bibfnamefont {J.}~\bibnamefont
  {Owrutsky}},\ }\bibfield  {title} {\bibinfo {title} {{Molecular polaritons
  for controlling chemistry with quantum optics}},\ }\href
  {https://doi.org/10.1063/1.5136320} {\bibfield  {journal} {\bibinfo
  {journal} {Journal of Chemical Physics}\ }\textbf {\bibinfo {volume} {152}},\
  \bibinfo {pages} {100902} (\bibinfo {year} {2020})}\BibitemShut {NoStop}%
\bibitem [{\citenamefont {Imperatore}\ \emph {et~al.}(2021)\citenamefont
  {Imperatore}, \citenamefont {Asbury},\ and\ \citenamefont
  {Giebink}}]{Imperatore2021}%
  \BibitemOpen
  \bibfield  {author} {\bibinfo {author} {\bibfnamefont {M.~V.}\ \bibnamefont
  {Imperatore}}, \bibinfo {author} {\bibfnamefont {J.~B.}\ \bibnamefont
  {Asbury}},\ and\ \bibinfo {author} {\bibfnamefont {N.~C.}\ \bibnamefont
  {Giebink}},\ }\bibfield  {title} {\bibinfo {title} {{Reproducibility of
  cavity-enhanced chemical reaction rates in the vibrational strong coupling
  regime Reproducibility of cavity-enhanced chemical reaction rates in the
  vibrational strong coupling regime}},\ }\href
  {https://doi.org/10.1063/5.0046307} {\bibfield  {journal} {\bibinfo
  {journal} {The Journal of Chemical Physics}\ }\textbf {\bibinfo {volume}
  {154}},\ \bibinfo {pages} {191103} (\bibinfo {year} {2021})}\BibitemShut
  {NoStop}%
\end{thebibliography}
\end{document}